\documentclass[onecollarge,natbib]{svjour2}
\bibpunct{[}{]}{;}{n}{}{,} 
\smartqed  
\usepackage{graphicx}
\journalname{Proceedings of Light Cone 2013}
\def\Journal#1#2#3{{#1} {\bf#2}: #3}

\def\NPB{{\rm Nucl. Phys.} B}

\def\PRD{{\rm Phys. Rev.} D}

\def\ep{\epsilon}

\def\la{\langle}
\def\ra{\rangle}

\def\lam{\lambda}

\def\be{\begin{equation}}
\def\ee{\end{equation}}
\def\bea{\begin{eqnarray}}
\def\eea{\end{eqnarray}}
\begin{document}

\title{Light-front zero-mode issue on the vector meson decay constant\thanks{This work was supported in part by Korea Research Foundation Grant (KRF-2010-0009019) and in part
by Kyungpook National University Research Fund, 2012.}
}


\author{Ho-Meoyng Choi         \and
        Chueng-Ryong Ji 
}


\institute{H.-M. Choi \at
              Department of Physics, Teachers College, Kyungpook National University, Daegu 702-701, Kore \\
              \email{homyoung@knu.ac.kr}           
           \and
           C.-R. Ji \at
            Department of Physics, North Carolina State University,
Raleigh, NC 27695-8202, U.S.A.\\
 \email{crji@ncsu.edu}
}

\date{Received: date / Accepted: date}

\maketitle

\begin{abstract}
We discuss the light-front zero-mode issue in the light-front quark model (LFQM) prediction
of a vector meson decay constant from the perspective of the vacuum fluctuation
consistent with the chiral symmetry of QCD.
We extend the exactly solvable manifestly covariant Bethe-Salpeter model calculation to the more phenomenologically accessible realistic LFQM and present a self-consistent covariant description of
the vector meson decay constant
analyzing the twist-2 and twist-3 quark-antiquark distribution amplitudes with even chirality.
\end{abstract}

\section{Introduction}
\label{intro}
Decay constants of mesons provide essential
information about the QCD
interaction between quark and antiquark.
They are the lowest moments of the light-cone distribution amplitudes (DAs) for the quark and antiquark inside the meson. Among various theoretical approaches to predict these fundamental constants of mesons,
the light-front quark model~(LFQM) based on the framework of light-front dynamics (LFD)~\cite{BPP} has become a very useful phenomenological tool to study various electroweak properties of mesons.
Owing to the rational energy-momentum dispersion relation, LFD has distinct features compared to other
forms of Hamiltonian dynamics. In particular, the feature that the vacuum fluctuations are suppressed
in LFD can be regarded as advantageous in hadron phenomenology.
However, one should also realize that the success of LFD in hadron physics cannot be realized unless the
treacherous points in LFD such as the zero-mode contributions in the hadron form factors and the
singularities associated with the rational energy-momentum relation are well taken care of with
proper methods. For instance, the zero-mode complication in the matrix element has been noticed
for the electroweak form factors involving a spin-0 and spin-1 particles as well as
the vector meson decay
constant~\cite{Jaus99,BCJ_spin1,CJ_Bc}.

Unlike the electroweak form factor described by a three-point function involving an external probe, the meson decay amplitude is described by a two-point function and may be regarded as one of the simplest possible physical
observables.
It is interesting that this apparently simple amplitude bears abundant fundamental
information on the QCD
vacuum dynamics and chiral symmetry.
To discuss the nature of the LF zero-mode in
the
meson decay amplitude,
we may denote the total LF longitudinal momentum of the meson, $P^+ = k_Q^+ + k_{\bar Q}^+$, where
$k_Q^+$ and $k_{\bar Q}^+$ are the individual quark and antiquark LF longitudinal momenta, respectively.
Similarly, the LF energy $P^-$ is shared by $k_Q^-$ and $k_{\bar Q}^-$, i.e. $P^- = k_Q^- + k_{\bar Q}^-$.
The LF energy integration is done typically by using
the Cauchy's theorem for a contour integration.
For the LF energy integration of the two-point function to compute the meson decay amplitude,
one may pick up a LF energy 
pole,
e.g. either $[k_Q^-]_{\rm on}$  (i.e. on shell value of $k_Q^-$)
from the quark propagator or $[k_{\bar Q}^-]_{\rm on}$ from the antiquark propagator.
However, it is crucial to note that the poles move to infinity (or fly away in the complex plane) as the LF longitudinal momentum, either $k_Q^+$ or $k_{\bar Q}^+$, goes to zero.
Unless the contribution from the pole flown into infinity vanishes, it must be kept in computing the physical observable. If such contribution exists,
since one of the constituents of the meson carries the entire momentum $P^+$ of the meson,
the other constituent carries the zero LF longitudinal momentum and thus can be regarded as the zero-mode quantum fluctuation linked to the vacuum.
It is important to capture the vacuum effect for the consistency with the chiral symmetry properties of the strong interactions.
With this link, the zero-mode contribution in the meson decay process can be considered effectively
as the effect of vacuum fluctuation consistent with the chiral symmetry of the strong interactions.
In this respect, the zero-mode issue of the vector meson decay constant in LFD is highly non-trivial
and deserves careful analyses.
It is a common practice to utilize an exactly solvable manifestly covariant model
to check the existence (or absence) of the zero-mode and substitute the radial and spin-orbit wave functions
of the exactly solvable model with the more phenomenologically accessible model wave functions that can be
provided by LFQM. Indeed, we found~\cite{CJ_fv13} that the existence or absence
of the zero mode may depend on the model, especially on the form of vector meson vertex operator,
while Jaus~\cite{Jaus99} claimed that there exists zero-mode contribution
even for the case of the plus component of the weak current $J^\mu_W$.

The purpose of this work is not only to clarify this zero-mode issue in the prediction of the decay
constant from LFQM but also to discuss this topic in relation to the vacuum fluctuation
consistent with the chiral symmetry of QCD.
To further clarify the zero-mode issue, we analyze the two-particle DAs and examine a
fundamental constraint anticipated from the LFQM~\cite{Jaus90,Cheng97,Kon,CJ_99}:
i.e. symmetric quark-antiquark DAs for the equal quark and antiquark bound state mesons such as $\rho$.
As we shall show in this work, the vector meson decay constant including the nonvanishing zero mode claimed by Jaus~\cite{Jaus99} and subsequently advocated by other authors~\cite{Cheng04} does not satisfy this anticipated constraint.
We also note that the two equivalent decay constants obtained from the plus component of the currents with the longitudinal polarization $(J^\mu_W,\epsilon_h)=(J^+_W,\epsilon_L)$ and the perpendicular
components of the currents with the transverse polarization
$(J^\perp_W,\epsilon_T)$ are related to the twist-2 and twist-3 two-particle DAs of
a vector meson according to the classification of Ball and Braun~\cite{Ball98,BJ07}, respectively.
With this aim, we extend our previous analysis~\cite{CJ_fv13}
to encompass the more phenomenologically accessible realistic LFQM~\cite{Jaus90,Cheng97,Kon,CJ_99}
and discuss a self-consistent covariant description of the vector meson decay constant in view of the link between QCD and LFQM.
More detailed analysis can be found in~\cite{CJ13}.

\section{Manifestly covariant model versus standard LFQM}
\label{sec:1}
The decay constant $f$ of a vector meson with the four-momentum $P$ and the mass $M$
as a $q{\bar q}$ bound state is defined by the matrix element of the vector
current

\be\label{eq:1}
\la 0|{\bar q}\gamma^\mu q|V(P,h)\ra
= f M \epsilon^\mu_h,
\ee
where 
$\epsilon_h$ is the polarization vector of a vector meson.

In the
manifestly covariant Bethe-Salpeter
model, the matrix element
${\cal J}^\mu_h \equiv \la 0|{\bar q}\gamma^\mu q|V(P,h)\ra$ is given in
the one-loop approximation as a momentum integral
\be\label{eq:2}
{\cal J}^\mu_h = N_c
\int\frac{d^4k}{(2\pi)^4} \frac{{\rm Tr}\left[\gamma^\mu\left(\slash \!\!\!p+m_1 \right)
 \Gamma\cdot\epsilon_h
 \left(-\slash \!\!\!k + m _2\right) \right]}
 {(p^2 -m^2_1 +i\varepsilon)(k^2 - m^2_2+i\varepsilon) } H_V,
\ee
where $N_c$ denotes the number of colors and $p =P -k$.
In order to regularize the covariant loop in $(3+1)$ dimensions,
we use $H_V = g/ N_\Lambda^2$ for the $q{\bar q}$ bound-state vertex function,
where $N_\Lambda  = p^2 - \Lambda^2 +i\varepsilon$, and $g$ and $\Lambda$ are constant parameters.
Although we use this particular dipole form of
the
hadronic vertex for our explicit analysis of covariant amplitude,
the essential features of our discussion linking the zero-mode quantum fluctuation to the vacuum consistent
with the chiral symmetry of QCD in corresponding LFD are independent of the specific form taken for the regularization
of the amplitude.
The manifestly covariant result $f^{\rm cov}$ is given by~\cite{CJ_fv13,CJ13}.
The meson vertex operator $\Gamma^\mu$ in the trace term is given by
$\Gamma^\mu =\gamma^\mu-(p-k)^\mu /D$.
While the Dirac coupling $\gamma^\mu$ is intrinsic to the vector meson vertex, the model-dependence of a
vector meson is implemented through the factor $D$.
For the explicit comparison between the manifestly covariant calculation and the LF calculation,
we analyze $\Gamma^\mu$ with a constant $D$ factor,
i.e. $D=D_{\rm con}=M + m_1 + m_2$.

%
%
Performing the LF energy $k^-$ integration of Eq.~(\ref{eq:2}),
we get the LF Bethe-Salpeter (LFBS) model of ${\cal J}^\mu_h$ as
\be\label{eq:14}
 [{\cal J}^\mu_h]_{\rm LFBS} = \frac{N_c}{16\pi^3}\int^{1}_0
 \frac{dx}{(1-x)} \int d^2{\bf k}_\perp
 \chi(x,{\bf k}_\perp) {\rm Tr}\left[\gamma^\mu\left(\slash \!\!\!p+m_1 \right)
 \Gamma\cdot\epsilon_h
 \left(-\slash \!\!\!k_{\rm on} + m _2\right) \right],
\ee
where
$\chi(x,{\bf k}_\perp) = g /[x^3 (M^2 -M^2_0)(M^2 - M^2_{\Lambda})^2]$
and $M^2_{0(\Lambda)} = [{\bf k}^{2}_\perp + m_1^2(\Lambda^2)]/x
 + ({\bf k}^{2}_\perp + m^2_2)/(1-x)$.
This result is obtained by closing the contour in the lower
half of the complex $k^-$ plane and picking up the residue at $k^-=k^-_{\rm on}$
 in the region $0< k^+ < P^+$ (or $0 < x < 1$).
To obtain the decay constant, we use two different combinations of the currents
and the polarization vectors,
i.e. (1) plus component  of the currents with the longitudinal
polarization (${\cal J}^+_L$)  and (2) perpendicular components
of the currents with the transverse polarization (${\cal J}^\perp_T$).

For the purpose of analyzing zero-mode contribution to the decay constant,
we denote the decay constant as $f_h^{\rm val}$ (meaning the valence
contribution to the decay constant)
when the matrix element ${\cal J}^\mu_h$ is obtained for $k^-=k^-_{\rm on}$ in the
region of $0<x<1$.
If $f_h^{\rm val}$ is different from the manifestly covariant result $f^{\rm cov}$,
this difference $f^{\rm cov}- f^{\rm val}_h$ corresponds to the zero-mode
contribution $f^{\rm z.m.}_h$ to the full solution
$f^{\rm full}_h=f^{\rm val}_h + f^{\rm z.m.}_h$.
The necessary prescription to identify zero-mode operator corresponding to
$p^-$ and the effective inclusion of the zero-mode operator in the valence region is
explicitly given in~\cite{CJ_fv13,CJ13}.
Both decay constants $f_{L}$ and $f_{T}$ obtained from ${\cal J}^{+}_{L}$ and
${\bf\cal J}^{\perp}_{T}$ receive the zero-mode contributions due to the singular $p^-$
(or equivalently $1/x$) term in the trace in the limit of $x\to 0$ when $p^-=p^-_{\rm on}$.
However, the source of the zero modes is different, i.e.  while $f_{L}$ receives the
zero mode from the model-dependent part including the $D_{\rm con}$ term, $f_T$ receives
the zero mode from the model independent Dirac coupling part, $\Gamma^\mu=\gamma^\mu$.
The explicit forms of $f^{\rm full}_{L(T)}$ can be found
in~\cite{CJ_fv13,CJ13}. We have checked that
$f^{\rm full}_T$ is the same as $f^{\rm full}_L$ as well as the
manifestly covariant result $f^{\rm cov}$.
We also confirmed that our $f^{\rm full}_T$ is exactly the same as the one
obtained by Jaus~\cite{Jaus99} (see Eq.~(4.22) of Ref.~\cite{Jaus99}).

On the other hand, in the standard LFQM presented in~\cite{Jaus90,Cheng97,Kon,CJ_99},
the constituent quark and
antiquark in a bound state are required to be on-mass-shell, which is different from the covariant
BS formalism in which the constituents are off-mass-shell. In the standard LF (SLF) approach used in
the LFQM~\cite{Jaus90,Cheng97,Kon,CJ_99},
the vector meson decay constant $f^{SLF}$ is obtained by the matrix element of the plus component of the
currents and the longitudinal polarization vector in 3-dimensional LF momentum space.
Effectively, the matrix element ${\cal J}^\mu_h$ in SLF is given by
\be\label{ApSLF}
[{\cal J}^\mu_h]_{\rm SLF} = \sqrt{N_c}\sum_{\lam_1\lam_2}\int\frac{dx d^2{\bf k}_\perp}{16\pi^3}
\phi(x,{\bf k}_\perp){\cal R}^{SS_z}_{\lam_1\lam_2}(x,{\bf k}_\perp)
\frac{\bar{v}_{\lam_2}(p_2)}{\sqrt{p^+_2}}\gamma^\mu\frac{u_{\lam_1}(p_1)}{\sqrt{p^+_1}},
\ee
where $\phi$ is the radial wave function and
${\cal R}^{SS_z}_{\lam_1\lam_2}=\bar{u}_{\lam_1}(p_1)\Gamma v_{\lam_2}( p_2)
/[2(M^{2}_{0}-(m_1-m_2)^{2})]^{1/2}$ is the
spin-orbit wave function that is obtained by the interaction-
independent Melosh transformation from the ordinary
spin-orbit wave function assigned by the quantum numbers
$J^{PC}$. The common feature of the standard LFQM
is to use the sum (i.e. invariant mass $M_0$) of the light-front energy of the constituent quark and antiquark for the meson mass in ${\cal R}^{SS_z}_{\lam_1\lam_2}$.
The virtue of using $M_0$
is to satisfy the normalization of ${\cal R}_{\lam_1 \lam_2}^{SS_z}$ automatically regardless of
any kinds of vector mesons, i.e.
$\sum_{\lam_1\lam_2}{\cal R}_{\lam_1 \lam_2}^{SS_z\dagger}{\cal R}_{\lam_1 \lam_2}^{SS_z}=1$.
The explicit helicity components of ${\cal R}_{\lam_1 \lam_2}^{SS_z}$ for a vector meson
are given in~\cite{CJ13}.
For the radial wave function $\phi$, we use the gaussian wave function
$\phi(x,{\bf k}_{\perp})=
(4\pi^{3/4}/\beta^{3/2}) \sqrt{\partial
k_z/\partial x} \;{\rm exp}(-{\vec k}^2/2\beta^2)$,
where $\beta$ is the variational parameter
fixed by the analysis of meson mass spectra~\cite{CJ_99} and
$\partial k_z/\partial x$ is the Jacobian of the variable transformation
$\{x,{\bf k}_\perp\}\to {\vec k}=({\bf k}_\perp, k_z)$.

\section{Correspondence between manifestly covariant model and standard LFQM}
\label{sec:2}
In the last section, we started from the manifestly covariant Bethe-Salpeter model given by Eq.~(\ref{eq:2})
and performed the LF energy integration to get the LFBS model given by Eq.~(\ref{eq:14}).
We also discussed the SLF model from the standard LFQM approach where the constituent quark and antiquark
are taken as the on-mass-shell degrees of freedom.
In this section, we shall analyze the relations between
$(f_L, f_T)$ obtained from $[{\cal J}^\mu_h]_{\rm LFBS}$
and $f^{SLF}$ from $[{\cal J}^\mu_h]_{\rm SLF}$.
If the on-mass-shell spin structure of the matrix element in SLF approach is exactly
reproducible from the manifestly covariant model, the SLF result is identified to be
Lorentz covariant.

For the direct comparison between $f_{(L,T)}$ and $f^{SLF}$, we find~\cite{CJ13} that
$f^{SLF}$ is exactly reproducible from the LFBS model
if and only if the following correspondence is applied
\be\label{eq:26}
 \sqrt{2N_c} \frac{ \chi(x,{\bf k}_\perp) } {1-x}
 \to  \frac{ \phi(x,{\bf k}_\perp) } {\sqrt{ x (1-x) [M^2_0 - (m_1-m_2)^2] }},\;\;
 M \to M_0.
 \ee
in the integrand of the formulae for $f^{\rm full}_L$ and $f^{\rm full}_T$.
The essential point of this replacement is to apply the replacement of $M\to M_0$ to
all physical mass terms in the integrands of $f^{\rm full}_L$ and $f^{\rm full}_T$.
With such self-consistent replacement, we find numerically that the three different forms
$f^{\rm full}_L$, $f^{\rm full}_T$ and $f^{SLF}_V$ indeed yield the identical result.
Moreover, we find that the on-shell contribution $f^{\rm on}_T$
to $f^{\rm full}_T$ also gives the same result with the other three, i.e.
$f^{\rm on}_T=f^{\rm full}_T=f^{\rm full}_L=f^{SLF}$ in the standard LFQM.
From those observations, we conclude that the replacement given by Eq.~(\ref{eq:26})
provides the self-consistent correspondence in connecting the
covariant BS model and the standard LFQM.
This self-consistent correspondence implies the proper absorption of the zero-mode contribution
essential for the link between chiral symmetry of QCD and LFQM.
More detailed discussions can be found
in~\cite{CJ13}.

Although those four different forms give the same result with each other when applying the
replacement in Eq.~(\ref{eq:26}), their quark DAs are quite different.
Therefore, by checking the DAs as an important constraint of the model,
we are able to further pindown the self-consistent LF covariant forms of the decay constant.
The quark DA of a vector meson, $\phi_V(x,\mu)$, is the probability of finding collinear quarks up to
the scale $\mu$ in the $L_z=0$ projection of the meson wave function.
The dependence on the scale $\mu$ is
given by the QCD evolution and can be calculated perturbatively.
However, the DAs at a certain low scale can be obtained by the necessary nonperturbative input
from LFQM. Moreover, the presence of the damping gaussian wave function allows us to perform
the integral up to infinity without loss of generality. The quark DA for a vector meson
is constrained by~\cite{CJ13}
\be\label{eq:DA}
\int^1_0\phi_V(x,\mu) dx = \frac{f_V}{2\sqrt{6}}.
\ee
One may also redefine the normalized quark DA as $\Phi_V(x)=(2\sqrt{6}/f_V)\phi_V(x)$ so that
$\int^1_0 dx \Phi_V(x)=1$.

For the equal quark and antiquark bound state meson such as $\rho$,
we find that only two forms of the decay constant, i.e.
$f^{SLF}$ from the longitudinal polarization and $f^{\rm on}_T$~\cite{CJ13}
(with the replacement given by Eq.~(\ref{eq:26})) from the transverse one,
yield the anticipated symmetric quark DA. The other two forms, i.e.
$f^{\rm full}_T$ and $f^{\rm full}_L$, that involve the corresponding zero-mode contributions
do not reproduce this fundamental constraint when $m_1 = m_2$.
The normalized quark DAs  obtained from $f^{SLF}$ and
 $f^{\rm on}_T$ (with the  replacement given Eq.~(\ref{eq:26})) correspond to the twist-2  $\phi^{||}_{2;V}(x)$
 and twist-3 $\phi^{\perp}_{3;V}(x)$, respectively.
 Our complete results for the twist-2 and twist-3 DAs in the standard LFQM are
 as follows~\cite{CJ13}:
\bea\label{T23DA}
 \phi^{||}_{2;V}(x) &=& \frac{2\sqrt{6}}{f_V}\int \frac{ d^2{\bf k}_\perp}{{16\pi^3}}
\frac{\phi(x,{\bf k}_\perp)}{\sqrt{{\bf k}^2_\perp + {\cal A}^2}}
\left[ {\cal A} + \frac{ 2 {\bf k}^2_\perp}{D_{\rm LF}} \right],
\\
\phi^{\perp}_{3;V}(x) &=& \frac{2\sqrt{6}}{f_V}\int \frac{ d^2{\bf k}_\perp}{{16\pi^3}}
\frac{\phi(x,{\bf k}_\perp)}{\sqrt{{\bf k}^2_\perp + {\cal A}^2}}
\frac{1}{M_0}\biggl\{ \frac{{\bf k}^2_\perp + {\cal A}^2}{2x(1-x)} -{\bf k}^2_\perp
  + \frac{ (m_1 + m_2)}{D_{\rm LF}}{\bf k}^2_\perp
 \biggr\},
\eea
where ${\cal A}= (1-x) m_1 + x m_2$ and $D_{\rm LF}=M_0 + m_1 + m_2$.
We note that while $f_V = f^{SLF}=f^{\rm on}_T$, $f_V$ used in $\phi^{||}_{2;V}(x)$
and $\phi^{\perp}_{3;V}(x)$ correspond
to $f^{SLF}$ (Eq.~(7)) and $f^{\rm on}_T$ (Eq.~(8)), respectively.

\section{Numerical Results}
\label{sec:Num}
In our numerical calculations within the standard LFQM, we use the set of the model parameters
for the harmonic oscillator confining potentials, i.e.
$(m_q, m_s, m_c)=(0.25,0.48,1.8)$ [GeV] and
$(\beta_{qq},\beta_{qs},\beta_{qc},\beta_{cc})=(0.3194,0.3419,0.4216,0.6998)$ [GeV] with $q=u$ or $d$,
which was obtained from the calculation of meson mass spectra using
the variational principle in our LFQM~\cite{CJ_99}.

\begin{table}[t]
\caption{Decay constants (in MeV) of $(\rho, K^*, D^*, J/\psi)$ mesons obtained
from our self-consistent covariant forms~\cite{CJ13} (see also Eqs.~(7) and (8)) compared with the experimental data~\cite{PDG12}. }
\centering
\label{tab:1}
\begin{tabular}{@{}ccccc}
\hline\noalign{\smallskip}
 & $f_\rho$ & $f_{K^*}$ & $f_{D^*}$ & $f_{J/\psi}$ \\ [3pt]
\tableheadseprule\noalign{\smallskip}
$f^{Th.}$ & 215 &  223 & 212 & 395 \\
$f^{EXP.}$ &
      220 (2)~\footnote{Exp. value for $\Gamma(\rho^0\to e^+ e^-)$.}, 209 (4)~\footnote{ Exp. value for $\Gamma(\tau\to\rho\nu_\tau)$.}
      & 217 (5) & - & 416 (6)\\
\noalign{\smallskip}\hline
\end{tabular}
\end{table}

%

In Table~\ref{tab:1}, we show the
results of the decay constants for ($\rho, K^*, D^*, J/\psi$) mesons obtained
from ($f^{\rm full}_L, f^{\rm on}_T, f^{\rm full}_T$) after applying the correspondence
in Eq.~(\ref{eq:26}) and $f^{SLF}$ and compare them with the experimental
data~\cite{PDG12}. We should note that the three different forms ($f^{\rm full}_L$,  $f^{\rm on}_T$,  $f^{\rm full}_T$) and $f^{SLF}$ are found~\cite{CJ13} to yield the same numerical
results $f^{Th.}$. While four different forms of the vector meson decay constant give
the identical results, they have different quark DAs.
Since $f^{\rm full}_L$ and $f^{\rm full}_T$ involve the corresponding zero-mode contributions,
they impose the intrinsic characteristic of the zero-modes, i.e. antisymmetric under
$x \leftrightarrow (1-x)$, inherited from the vacuum property~\cite{CJ13}. Thus,
the quark DAs from $f^{\rm full}_L$ and $f^{\rm full}_T$ do not satisfy the expected
constraint, i.e. symmetric DAs for the meson with $m_1 = m_2$ such as $\rho$.
However, $f^{SLF}$  and $f^{\rm on}_T$, which is free from the explicit
zero-mode contribution, yield the anticipated symmetric DAs for the meson with $m_1 = m_2$
and thus provide self-consistent LF covariant descriptions of vector
meson decay constants in the standard LFQM.
Since  $f^{SLF}$ [$f^{\rm on}_T$] is obtained from using ($J^+_W, \ep_L$)
[ ($J^\perp_W, \ep_T$) ], the normalized quark DAs  obtained from $f^{SLF}$ and
$f^{\rm on}_T$  correspond to the twist-2 DA $\phi^{||}_{2;V}(x)$ (Eq.~(7)) and
twist-3  DA $\phi^{\perp}_{3;V}(x)$ (Eq.~(8)), respectively.

\begin{figure}
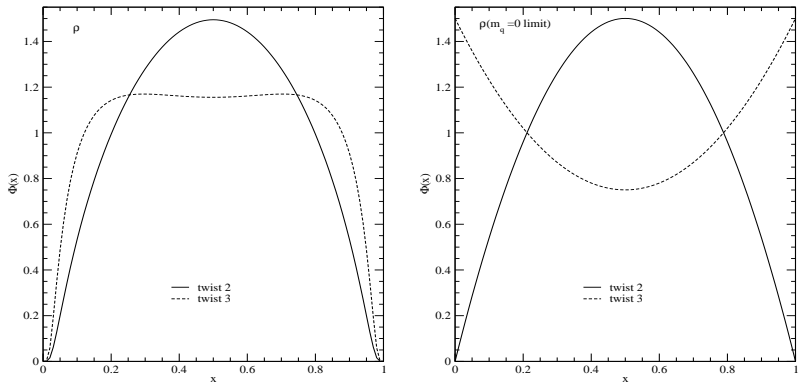

\vspace{0.5cm}
\centering
\includegraphics[height=5cm, width=5cm]{Dfig3a.eps}
\hspace{0.2cm}
\includegraphics[height=5cm, width=5cm]{Dfig3b.eps}
\caption{ The twist-2 DAs $\phi^{||}_{2;V}(x)$ and twist-3
DAs $\phi^{\perp}_{3;V}(x)$ for $\rho$ meson with nonzero constituent quark masses
(left panel) compared to those with massless quark case (right panel)~\cite{CJ13}.
}
\label{fig:1}
\end{figure}

In Fig.~\ref{fig:1}, we show  $\phi^{||}_{2;V}(x)$ (solid line) and $\phi^{\perp}_{3;V}(x)$ (dashed line) of the  $\rho$ meson in cases of  finite constituent quark masses
 (left panel)  and massless quark respecting chiral symmetry (right panel).
 We should note that our LFQM predictions of twist-2 and twist-3 DAs in the chiral
symmetry ($m_q\to 0$) limit remarkably reproduce the exact asymptotic DAs, i.e.
$[\phi^{||}_{2;V}(x)]_{\rm as}=6 x (1-x)$ and $[\phi^{\perp}_{3;V}(x)]_{\rm as}=(3/4)(1+\xi^2)$
where $\xi=2x -1$ anticipated from QCD sum rule predictions~\cite{Ball98}.
This example may show again that our LFQM prediction satisfies the chiral symmetry consistent with
the QCD if one correctly implement the zero-mode link to the QCD vacuum.
The quark mass correction is not large for the twist-2 $\phi^{||}_{2;V}(x)$,  however,
it is very significant for the case of twist-3 $\phi^{\perp}_{3;V}(x)$ especially
at the end point regions.

\section{Summary}
\label{sec:sum}

In this work, we extended our previous analysis~\cite{CJ_fv13} of the vector meson decay constant from the exactly
solvable manifestly covariant BS model to the more phenomenologically accessible realistic LFQM~\cite{Jaus90,Cheng97,Kon,CJ_99}. We discussed a
self-consistent covariant description of the vector meson decay constant in view of the
link between the chiral symmetry of QCD and the expected numerical results of the LFQM.

As the SLF approach within the LFQM by itself is not amenable to determine the zero-mode contribution, we utilized the manifestly covariant model to check the existence (or absence) of the zero-mode.
Performing a LF calculation in the covariant BS model, we computed the decay constants using two different combinations of
LF weak currents $J^\mu_W$ and polarization vectors $\epsilon_h$, i.e. $f_{h=L}$ obtained from
$(J^\mu_W,\epsilon_h)=(J^+_W,\epsilon_L)$ and $f_{h=T}$ from $(J^\perp_W,\epsilon_T)$,
 and checked the LF covariance of the decay constants.
We found in the manifestly covariant model that both combinations gave the same result with some
particular LF vertex functions if  the missing zero-mode contributions were properly taken into account.
We then substituted the radial and spin-orbit wave functions
with the phenomenologically accessible model wave functions provided by the LFQM and
compared $f_{(L,T)}$  obtained from the BS model with the decay constant $f^{SLF}$ obtained
directly from the SLF approach used in the LFQM~\cite{Jaus90,Cheng97,Kon,CJ_99}.
Linking the covariant BS model to the standard LFQM, we found the matching condition
given by Eq.~(\ref{eq:26}) between the two to give a self-consistent covariant description
of the decay constant within the LFQM.
Using that correspondence, we were able to pin down two independent covariant forms of vector meson decay constants,
one obtained from $(J^+_W,\epsilon_L)$ and the other from $(J^\perp_W,\epsilon_T)$.
Although both of them yield the identical decay constant, each of them corresponds to different twist DA:
$(J^+_W,\epsilon_L)$ and $(J^\perp_W,\epsilon_T)$ correspond to twist-2 and twist-3 two-particle DAs, respectively.
Our twist-2 and twist-3 DAs not only satisfy  the fundamental constraint of the DAs anticipated from the isospin symmetry, i.e.
symmetric DAs for the equal quark and antiquark bound state mesons (e.g. $\rho$ meson), but also reproduce the correct asymptotic DAs in the chiral
symmetry limit.  Further analysis including the chirality odd and higher twist DAs is under consideration.




\end{document}